\documentstyle[epsfig]{aipproc}
\begin{document}

\title{{A High Peculiarity Rate for Type Ia SNe}}

\author{W. D. Li$^*$, A. V. Filippenko$^*$,  A. G. Riess$^*$\thanks{Now 
at Space Telescope Science Institute, Baltimore, MD 21218},\\
J. Y. Hu$^{\dagger}$, and Y. L. Qiu$^{\dagger}$} 
\address{$^*$Department of Astronomy,
University of California, Berkeley, CA 94720-3411 USA\\
email: (wli, alex)@astro.berkeley.edu\\
$^{\dagger}$Beijing Astronomical Observatory, Chinese Academy of Sciences, Beijing 100080
 China}

\maketitle

\section*{ INTRODUCTION}

Type Ia supernovae (SNe Ia) are not perfectly homogeneous. There are peculiar
ones: SN 1991T-like (overluminous), SN 1986G-like (subluminous), and SN
1991bg-like (very subluminous) objects. Figure 1 shows a comparison of the
spectra of peculiar SNe~Ia with that of a relatively normal SN Ia, SN 1994D.

The peculiarity rate, however, is not well established.  An estimate (less than
10\%) by Branch et al. (1993) [1] is limited by the small number of peculiar
SNe~Ia known at that time.

In the past 3 years, a number of peculiar SNe have been discovered in the
course of several successful nearby SN surveys, and we try to update the
peculiarity rate here.

\begin{figure}[b!]
\centerline{\epsfig{file=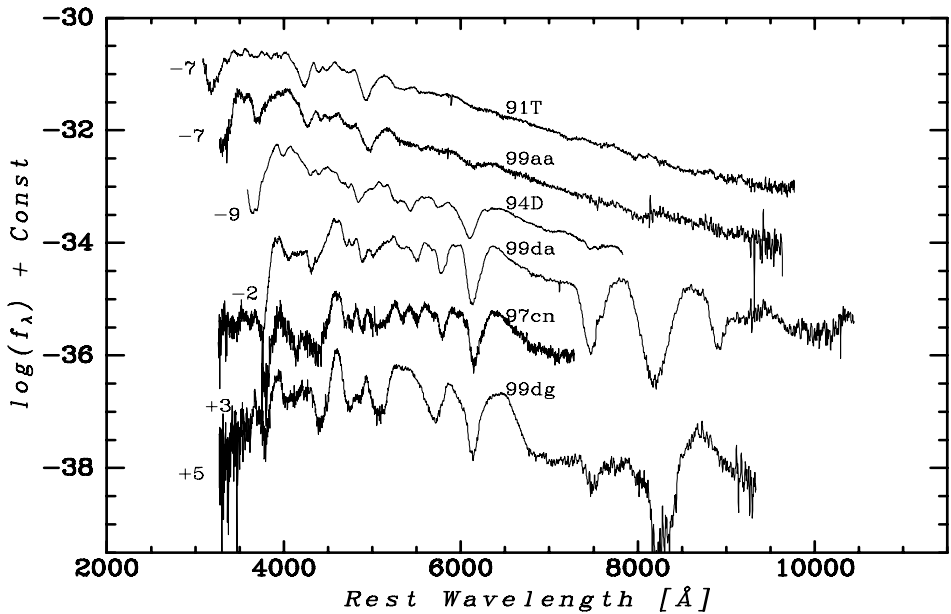,height=3.5in,width=5.5in}}
\vspace*{10pt}
\caption{ Spectra of two SN 1991T-like objects (99aa, 91T), one normal object
(94D), and three SN 1991bg-like objects (99da, 97cn, 99dg). Days are
relative to maximum.}
\label{fig1}
\end{figure}

\section*{ THE SN Ia SAMPLE}

   We have compiled a sample of 90 SNe Ia from 1997 to 1999 (up to SN 
1999da). The only criterion we used to select SNe~Ia in the sample is that 
the redshift of the SN must be smaller than 0.1. This is to avoid
high-redshift SNe~Ia to ensure that there is no contamination from 
possible evolution and/or observational biases between nearby and 
high-redshift SNe~Ia.

   The SNe Ia in our sample are subclassified as normal or as one of 
the peculiar SNe~Ia: SN 1991T, which had prominent Fe~III absorption
lines and weak Si~II lines prior to and near maximum brightness [2, 3]; 
SN 1991bg, which had an enhanced Si~II 5700\AA\ absorption, and a broad 
absorption trough extending from about 4150 to 4400\AA\ due to 
Ti II lines [4, 5]; SN 1986G, which also had Ti absorption but
less prominent than in SN 1991bg [6, 7]. Classification is done 
based on information in the International Astronomical Union 
Circulars (IAUC) and our SN spectrum and photometry database.

\section*{ THE OBSERVED PECULIARITY RATE}

We have divided our SN~Ia sample into several subsamples and reported the
observed peculiarity rates as follows.

In the total sample, all 90 SNe are considered. 17 (18.9\%) SNe are peculiar, among
which 11 (12.2\%) SNe are SN 1991T-like and 6 (6.7\%) SNe are SN 1991bg/1986G-like.

   In the near-maximum sample, only the SNe Ia that were
spectroscopically classified at no later than a week after
maximum are considered. 61 SNe are in the sample, among which 17 (27.9\%)
are peculiar. 11 (18.0\%) SNe are SN 1991T-like and 6 (9.9\%) SNe
are SN 1991bg/1986G-like.

   In the Lick-Beijing (LB) sample, only the SNe that were discovered in
the sample galaxies of the Lick Observatory Supernova
Search (LOSS) and the Beijing Astronomical Observatory
Supernova Survey (BAOSS) are considered. 35 SNe are in 
the sample, among which 13 (37.1\%) are peculiar.  7 (20.0\%) 
SNe are SN 1991T-like and  6 (17.1\%) SNe are SN 1991bg/1986G-like.

\begin{figure}[b!]
\centerline{\epsfig{file=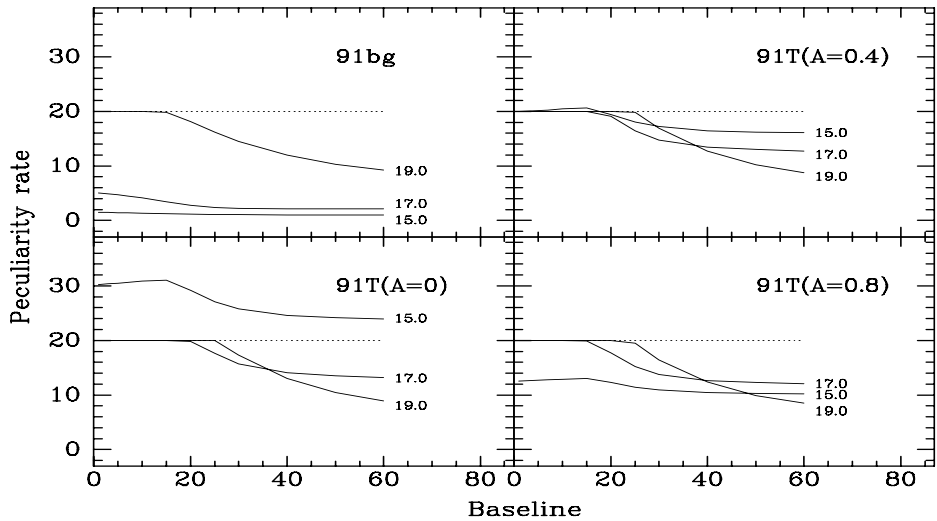,height=3.0in,width=5.5in}}
\vspace*{10pt}
\caption{The peculiarity rate found in the Monte Carlo simulations for 
distance-limited SN surveys. Results are shown for different baselines
and different limiting magnitudes, and for the case of SN 1991bg-like
objects (upper left), SN 1991T-like objects with no extra extinction
(lower left), SN 1991T-like objects with 0.4 mag of extra extinction
(upper right), and SN 1991T-like objects with 0.8 mag of extra 
extinction (lower right). The dotted line is the input (intrinsic) rate.
}
\label{fig2}
\end{figure}

\section*{OBSERVATIONAL BIASES}

There are various observational biases that make the observed peculiarity rate
deviate from its intrinsic value.

\begin{enumerate}

  \item{{ The maximum-only bias} -- caused by the fact that SN 1991T-like
  objects can best be spectroscopically identified prior to or near maximum
  brightness. It is thus unknown whether a SN discovered well after maximum is
  normal or SN 1991T-like. Classifying them as normal causes the maximum-only
  bias that underestimates the peculiarity rate. }

  \item{{ The Malmquist bias } -- caused by the difference in luminosity among
  SNe Ia.}

  \item{ { The light-curve shape bias} -- caused by the difference in
light-curve shape among SNe Ia.}

  \end{enumerate}

\section*{ MONTE CARLO SIMULATIONS}

  We have done Monte Carlo simulations to study the effects of the
observational biases. Simulations are done for magnitude-limited SN surveys,
with the baseline as the only input parameter.  Simulations are also done for
distance-limited SN surveys, with the baseline and the limiting magnitude of
the survey as parameters.

 All observational biases are well accounted for in the Monte Carlo simulations.
We also studied the role of extinction of SN 1991T-like objects in determining
the rate of those objects. There are speculations that SN 1991T-like objects
are more likely to occur in dusty, star-forming regions and thus may suffer
more extinction than the normal and SN 1991bg/1986G-like objects.

 An example of the results of the Monte Carlo simulations is shown in Figure 2.


\section*{ THE INTRINSIC PECULIARITY RATE}

Our simulations indicate that {\bf all} SNe~Ia in the LB sample should have
been discovered because the two surveys are distance-limited with small
baselines and deep limiting magnitudes. In other words, the peculiarity rate
and luminosity function in the LB sample should be intrinsic.

  Our simulations also indicate that a high peculiarity rate (more than 30\%)
and a flat luminosity function for SN~Ia (e.g., the rate of SN 1991bg-like and
SN 1991T-like objects are comparable) are consistent with the observed
peculiarity rates in all three (total, near-maximum, and LB) samples.

  These results have important implications for studies of high-redshift 
SNe~Ia and of SN~Ia progenitor systems.

\begin{enumerate}

  \item{The high-redshift results: The high peculiarity rate for nearby SNe~Ia
  ($>$30\%) and the small apparent (but preliminary) peculiarity rate for the
  several dozen high-redshift SNe~Ia studied thus far [8, 9] may indicate a
  systematic difference between them. To reconcile the absence of SN 1991T-like
  objects found at high redshifts with the peculiarity rate found at low
  redshifts, an extra extinction of more than $\sim 1$ mag for the SN 199T-like
  objects is needed, which is not supported by the observations. However, the
  existing spectral studies of high-redshift SNe~Ia are not very detailed.}

  \item{The progenitor systems of SNe~Ia: The high peculiarity rate, together
  with other evidence, favors the existence of multiple progenitor systems for
  SNe~Ia (e.g., single-degenerate systems and double-degenerate systems).}

  \end{enumerate}

\bigskip
  Our supernova research at UC Berkeley is supported by NSF grant
AST-9417213 and NASA grant GO-7434.

\end{document}